\documentclass[aps,pra,reprint,superscriptaddress]{revtex4-1}

\usepackage{graphicx}

\usepackage{bm}

\pdfoutput=1

\begin{document}

\title{Compensating fictitious magnetic field gradients in optical microtraps by using elliptically polarized dipole light}

\author{S\'{e}bastien Garcia}
\email[]{sgarcia@phys.ethz.ch}

\affiliation{Laboratoire Kastler Brossel,  ENS-PSL Research University, CNRS, UPMC-Sorbonne Universit\'{e}s, Coll\`{e}ge de France, 24 rue Lhomond, 75005 Paris, France}

\affiliation{Department of Physics, ETH Z\"urich, CH-8093 Z\"urich, Switzerland}

\author{Jakob Reichel}

\affiliation{Laboratoire Kastler Brossel,  ENS-PSL Research University, CNRS, UPMC-Sorbonne Universit\'{e}s, Coll\`{e}ge de France, 24 rue Lhomond, 75005 Paris, France}

\author{Romain Long}
\email[]{long@lkb.ens.fr}

\affiliation{Laboratoire Kastler Brossel,  ENS-PSL Research University, CNRS, UPMC-Sorbonne Universit\'{e}s, Coll\`{e}ge de France, 24 rue Lhomond, 75005 Paris, France}

\date{\today}

\begin{abstract}
  Tightly focused optical dipole traps induce vector light shifts
  (``fictitious magnetic fields'') which complicate their use for
  single-atom trapping and manipulation. The problem can be mitigated
  by adding a larger, real magnetic field, but this solution is not
  always applicable; in particular, it precludes fast switching to a
  field-free configuration. Here we show that this issue can be
  addressed elegantly by deliberately adding a small elliptical
  polarization component to the dipole beam. In our experiments with
  single $^{87}$Rb atoms in a chopped trap, we observe improvements up
  to a factor 11 of the trap lifetime compared to the standard,
  seemingly ideal linear polarization. This effect results from a modification of heating processes via spin-state diffusion in state-dependent trapping potentials. We develop Monte-Carlo simulations of the evolution of
  the atom's internal and motional states and find that they agree
  quantitatively with the experimental data. The method is general and
  can be applied in all experiments where the longitudinal
  polarization component is non-negligible.

\end{abstract}


\maketitle

\section{\label{sec:intro}Introduction}

When a linearly polarized electromagnetic beam is focussed so tightly
that the paraxial approximation breaks down, its polarization near the
focus is no longer purely transverse, but acquires a longitudinal
component that increases with the confinement of the field and leads
to polarization gradients in the strongly confined region. The effect
has been known for a long time \cite{Richards1959}, but currently gains
importance in the context of light-matter interfaces with ultracold
atoms. While similar effects play an enabling role in the new field of
chiral quantum optics \cite{lodahl2016chiral}, they complicate the
use of strongly focussed, far-detuned laser beams for trapping and
manipulating single atoms. Such single-atom tweezers
\cite{Schlosser2001singleatom} are a powerful tool with many
applications. They have been used, for example, to prepare a single
trapped atom by collisional blockade \cite{Schlosser2001singleatom},
to entangle single atoms by Rydberg blockade \cite{Urban2009,Gaetan2009},
to laser cool a single atom to its vibrational ground state
\cite{Kaufman2012, Thompson2013}, and to couple an atom to a near-field optical
structure \cite{Thompson2013b}. When realized with a
pigtailed optical fiber, a single-atom tweezer also enables an
attractively simple source of narrowband single photons
\cite{garcia2013fiber}.

For standard dipole traps, the polarization is commonly chosen to be
linear to ensure an equal light shift for all magnetic sublevels of
the ground state \cite{Grimm2010dipole}. In single-atom tweezer
experiments, where tight focussing is mandatory, the atom experiences
an elliptical polarization with spatial variations on the wavelength
scale even for a linearly polarized input field.  The result is a
vector light shift, which depends on the magnetic quantum number and
can be described by a an effective, ``fictitious'' magnetic field
\cite{Cohen1972fictif}. One approach to mitigate this effect is to add a
stronger, real magnetic field in a direction orthogonal to the
fictitious field in order to reduce the field gradient
\cite{Kaufman2012,Thompson2013}. It is however not always possible to
add strong real magnetic fields, whose switching dynamics are usually
slow.

In this article, we experimentally demonstrate that a properly chosen
non-linear input polarization can be used instead of a real magnetic
field to mitigate the damaging effects of vector light shift
gradients. Our single-atom tweezer is chopped at rather high
frequencies up to 4\,MHz. In this way, spectroscopy, optical cooling and single-photon
generation can be performed on an unperturbed atom during the
field-free periods, while still providing a strong effective trapping
potential \cite{Chu1986, garcia2013fiber, Hutzler2017}.  As we show, a slightly elliptical polarization for the input field of the chopped tweezer can improve the lifetime of the
trapped atom by more than an order of magnitude. This is against the
common belief that linear polarization is the best choice for an
optical dipole trap, which actually holds only as long as the
longitudinal component of the electromagnetic field can be
neglected. We analyze how the experimental lifetime of the trapped
atom depends on the trap chopping frequency and on the dipole light
polarization. We then develop a model based on trap modifications due
to quantum jumps between different Zeeman sub-states of the atom and
perform Monte-Carlo simulations in very good agreement with the
experimental data.

\section{\label{sec:exp}Measurement of single atom lifetime for different polarizations}

\begin{figure}[tb]
\includegraphics[width = 0.95\columnwidth]{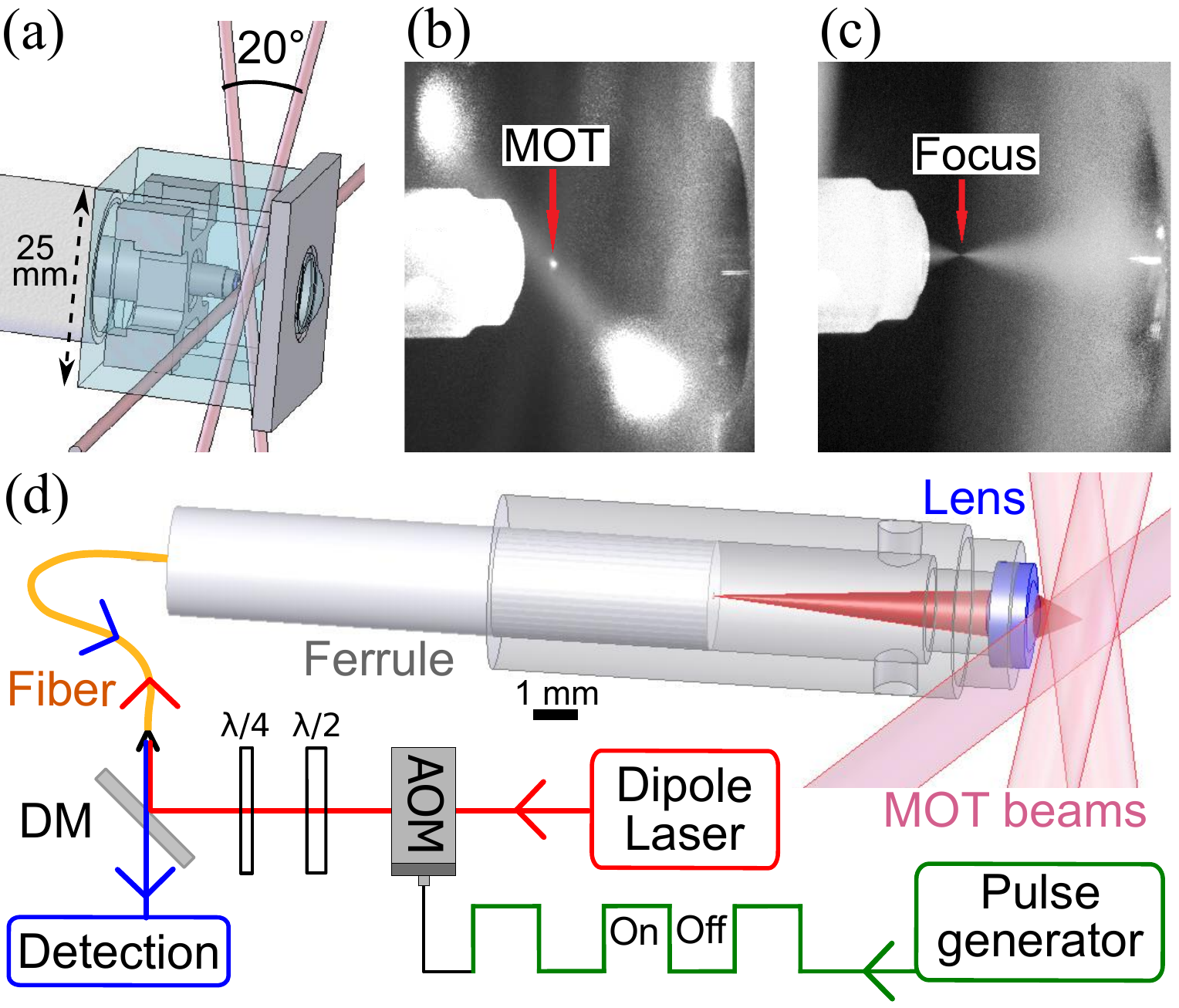}
\caption{(a) Experimental setup:  the fiber-pigtailed optical tweezer is placed inside a cubic 25$\,$mm size all-glass cell. To load the dipole trap, we produce a cloud of laser-cooled atoms in a MOT with three retro-reflected 1$\,mm$ diameter beams, two of them crossing at an angle of 20 deg to avoid clipping at the lens.
(b) Image by fluorescence of a $^{87}$Rb atomic cloud in the MOT. For this image, the rubidium pressure inside the cell has been strongly increased to make the cloud visible.
(c) With the same high-level pressure and by sending resonant light at 780 $\,$nm in the fiber, we can see the position of the tweezer focus by imaging the fluorescence of the vapor background atoms. As the focus at 810 $\,$nm is 2$\,\mu$m apart, the dipole light is well focussed at the center of the MOT.
(d) The optical tweezer collects also the resonant photons from  trapped single atoms. An acousto-optic modulator (AOM) chops the dipole laser whose polarization is controlled by two waveplates.}\label{fig:setup}
\end{figure}

In our experimental setup (see Fig.~\ref{fig:setup}) described
in detail in Ref.~\cite{garcia2013fiber}, we have developed a
miniature and robust fiber-pigtailed optical tweezer, where the same
fiber is used to trap a single $^{87}$Rb atom and to read-out its
fluorescence.  We have also demonstrated that, once loaded, the
pigtailed tweezer can be used as a single-photon source at 780$\,$nm.
To load the tweezer, we produce a laser-cooled atoms cloud in a
magneto-optical trap (MOT) at the focus of a far off-resonance optical
dipole trap with a wavelength close to 810$\,$nm and a measured waist
of $1.4\,\mu$m. To eliminate trap-induced light shifts, which reduce
the efficiency of the MOT cooling, constitute a source of spectral
broadening, and to avoid the generation of 780$\,$nm photons by
anti-Stokes Raman scattering of the trapping light inside the fiber,
we chop the dipole light with an acousto-optical modulator 
(see Fig.~\ref{fig:setup}.d) . If the chopping frequency is much larger than twice the largest trap
oscillation frequency, the atom is expected to stay trapped in this
attractive time-averaged potential. When the dipole light is on, the trap depth is $5.6\,$mK, and the calculated transverse and longitudinal trap frequencies are $f_{\mathrm{t}} \simeq 167\,$kHz and $f_{\mathrm{l}} \simeq 22\,$kHz.

\begin{figure}[tb]
\includegraphics[width = 0.95\columnwidth]{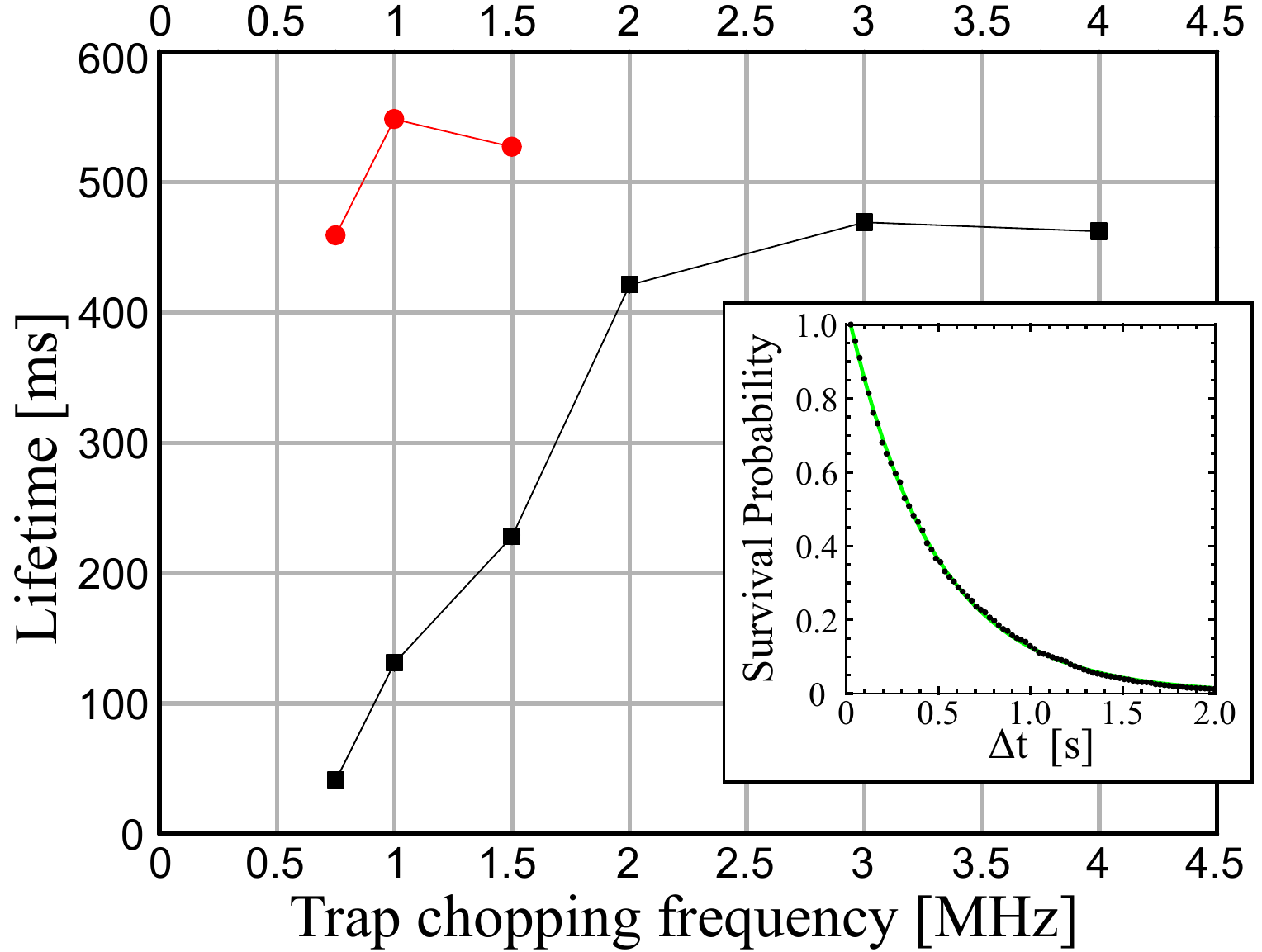}
\caption{Observed lifetimes as function of trap modulation frequency for linear polarized dipole light (black squares) and for an optimized elliptic polarization with $\alpha_{\mathrm{opt}}  \simeq 7\,$deg  correponding to $\vert C_z \vert \simeq 0.24$ (red dots). Inset: example of lifetime extraction by exponential fit (green) of experimental survival probability decay (black dots) as function of time ($\Delta t$) elapsed since the atom entered in the trap, for a linear polarization with $3$ MHz chopping frequency.}\label{fig:data}
\end{figure}

We have investigated the effect of the chopping frequency on the lifetime of the atom inside the trap (see Fig.~\ref{fig:data}) in the presence of the MOT beams used to load the dipole trap. The lifetime is obtained by an exponential fit of the survival probability of the atom versus the duration of the trapping. To be mainly limited by background gas collisions, we work in the weak loading regime, where loss due to a second atom entering the trap is negligible.  
For linearly polarized dipole light, the lifetime was limited by background gas collisions  at a value of about $500\,$ms for chopping frequencies above $3\,$MHz. Below $3\,$MHz, the lifetime decreases with the chopping frequencies and no single-atom trapping signal could be detected below 500$\,$kHz. So, with this polarization, the chopping frequency needs to be much larger, by about one order of magnitude, than the largest calculated trap frequency.
For chopping frequencies between $750\,$kHz and $1.5\,$MHz, we observe a strong improvement of the lifetime by adding a small circular component to the polarization of the dipole light. The lifetime is found optimal for an elliptical polarization characterized by the angle $\alpha_{\mathrm{opt}} = \arctan(E_b/E_a) \simeq 7\,$deg, where $E_a$ and $E_b$ are the semi-major and semi-minor axes of the electric-field ellipse, respectively. At a chopping frequency of $750\,$kHz, the lifetime is improved by a factor $11$.

These results cannot be explained by the chopping of the dipole light, which  is expected to be negligible when the trap chopping frequency $f_{\mathrm{m}}$ becomes higher than the largest parametric heating frequency ($2 f_{\mathrm{t}}$) of the trap~\cite{savard1997laser}. Moreover, this effect applies to all polarization configurations and so cannot explain the improved lifetime observed for slightly elliptical polarization. 
So, we can deduce from these results that additional heating mechanisms reduce the lifetime for a linearly dipole light and that this effect can be compensated by using an elliptical polarization.

\section{\label{sec:sublevels}Effect of the polarization on the optical traps for the different Zeeman sub-levels}

The modification of the lifetime induced by polarization finds its
origin in the dipole potential of the ground state which depends on
the Zeeman sub-level the atom occupies. For a dipole-light frequency
close to the D1 and D2 transition frequencies of an alkali atom, this
potential is given by~\cite{kuppens2000loading,Thompson2013}:
\begin{equation}
U_{\mathrm{dip}}\left(\bm{r}\right) = U_{\mathrm{dip},0}\left(\bm{r}\right)  \left( 1 + \frac{\delta_1 - \delta_2}{2\delta_1 + \delta_2} \bm{C} \left( \bm{r} \right) .\frac{g_F}{\hbar} \hat{ \bm{F} } \right) 
\label{eq:Udip}
\end{equation}
\begin{equation}
\mathrm{with}  \ \ \  U_{\mathrm{dip},0}\left(\bm{r}\right) = \frac{\hbar \Gamma }{24 } \left( \frac{\Gamma}{\delta_{1}} + 2 \frac{\Gamma}{\delta_{2}} \right) \frac{I\left(\bm{r}\right)}{I_{\mathrm{S}}}
\label{eq:Udip0}
\end{equation}
the scalar dipole potential. In these equations, $\Gamma$ is the
natural linewidth of the excited P levels,
$\delta_i = \omega_{\mathrm{L}} - \omega_i$ are the detunings between
the laser and the D1 and D2 transition frequencies ($i={1,2}$
respectively), $I\left(\bm{r}\right)$ the light intensity and
$I_{\mathrm{S}}$ the saturation intensity. We note $\hat{\bm{F}}$ the
atom angular momentum operator (norm $F=2$ here) and
$g_F = [F(F+1) - I(I+1) +J(J+1)]/[F(F+1)]$.
$\bm{C}\left(\bm{r}\right) =
\Im\left(\bm{e}_{\mathrm{p}}\left(\bm{r}\right) \times
  \bm{e}_{\mathrm{p}}\left(\bm{r}\right)^* \right)$ is a vector
characterizing the ellipticity and direction of the polarization
represented by a unit vector
$\bm{e}_{\mathrm{p}}\left(\bm{r}\right)$. For a polarization
$\bm{e}_{\mathrm{p}} = i \sin(\alpha) \bm{e}_x + \cos(\alpha)
\bm{e}_y$, we have
$\bm{C} = \sin(2\alpha) \bm{e}_z$: for linear polarization
$\bm{C} = \bm{0}$, and for a circular polarization $\Vert\bm{C}\Vert = 1$. When
$\bm{C}$ is non-zero, the different $m_F$ sub-levels
experience different light-shifts that are analogous to energy shifts
induced by a fictitious magnetic field given by:
\begin{equation}
\bm{B_{\mathrm{fict}}}\left(\bm{r}\right) = - U_{\mathrm{dip},0}\left(\bm{r}\right)  \frac{|\delta_1 - \delta_2|}{2\delta_1 + \delta_2} \frac{g_F}{\mu_{\mathrm{B}} g_{\mathrm{L},F} } \bm{C}\left(\bm{r}\right),
\label{eq:Bfict}
\end{equation}
where $\mu_{\mathrm{B}}$ is the Bohr magneton and $g_{\mathrm{L},F}$ is the Land\'{e} g-factor, which is equal to $g_F$ in the usual limits where the electron spin, electron orbital and nuclear Land\'{e} g-factors are approximated to $g_{\mathrm{L},S} \simeq 2$, $g_{\mathrm{L},L} \simeq 1$ and $g_{\mathrm{L},I} \simeq 0$, respectively. 

\begin{figure}[t]
\includegraphics[width = 0.95\columnwidth]{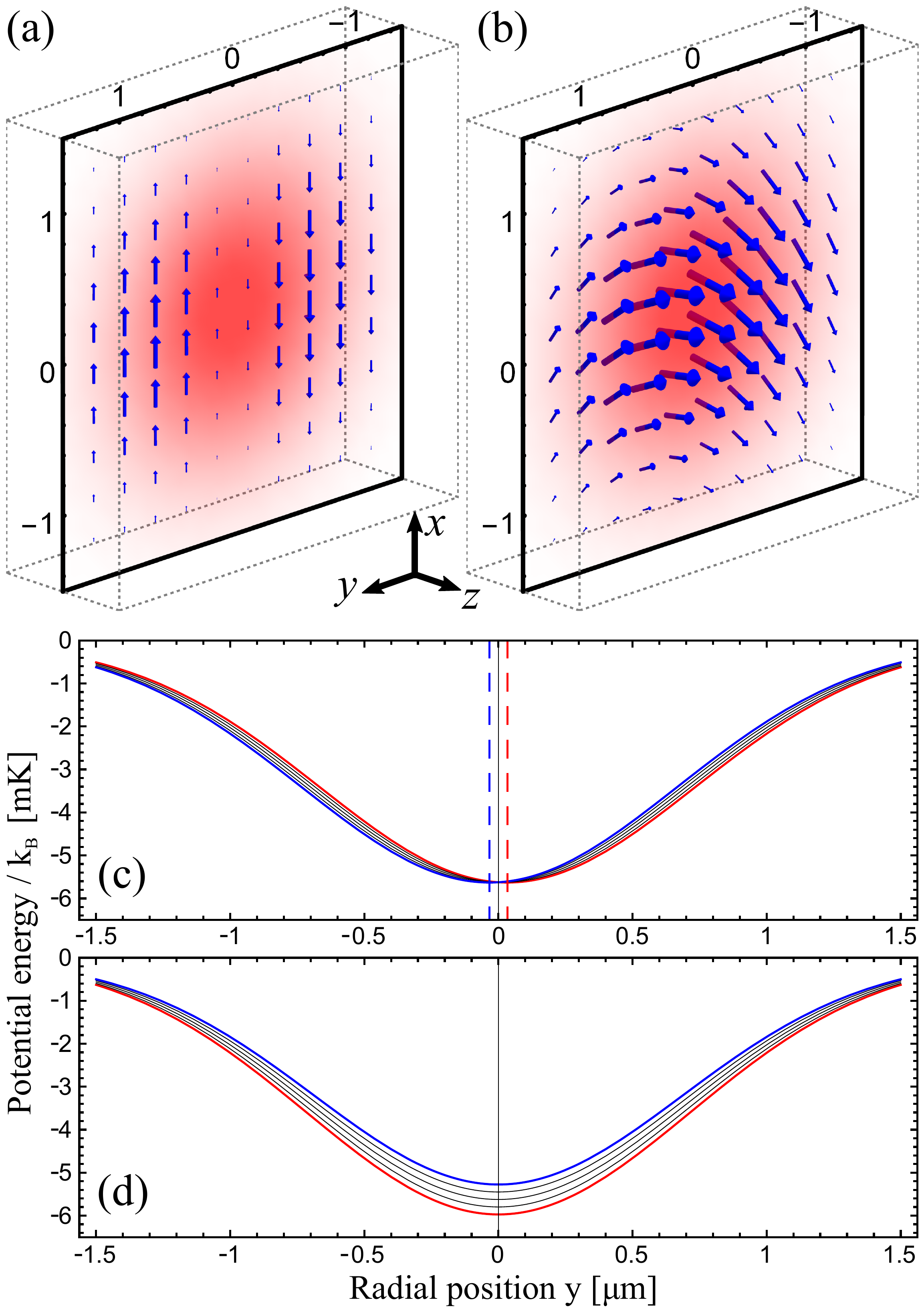}
\caption{Dipole potential in the fiber-pigtailed tweezer: (a)
  Representation of the vector $\bm{C}\left(\bm{r}\right) \vert U_{\mathrm{dip},0}\left(\bm{r}\right) \vert$, which is opposite of  the fictitious  magnetic field $\bm{B_{\mathrm{fict}}}$ by blue vectors   in the $xy$
  focal plane (the beam propagates towards increasing values of $z$)
  for a linear polarization aligned with the $y$-axis. The red density plot represents the scalar  		dipole potential. (b) Same as in (a) for optimal-lifetime elliptic
  polarization. (c) Dipole potentials of the five Zeeman sub-levels
  from $m_F=-2$ (red) to $m_F = 2$ (blue) of the $F=2$ ground
  state for a linear polarization aligned with the $y$-axis. The traps
  are mainly displaced by a regular interval (the trap centers of the
  extreme states are indicated by dashed vertical lines). (d) Same as
  in (c) for the case of the optimal-lifetime elliptic
  polarization. In this case, the trap depths and the oscillation
  frequencies are modified.}\label{fig:fieldpot}
\end{figure}

In our optical tweezer, a Gaussian beam is strongly focussed down to a waist $w_0 \simeq 1.4 \,\mu$m. For such a diverging beam, a purely transverse incident light (which is linearly-polarized along the $y$-axis)  generates at the focus plane strong variations of the local polarization due to interferences of the different focused light rays. The local $\bm{C}\left(\bm{r}\right)$ characteristic vector is calculated via vector Debye integral~\cite{Richards1959} and we plot the product $\bm{C}\left(\bm{r}\right) \vert U_{\mathrm{dip},0}\left(\bm{r}\right) \vert$, which is opposite of the fictitious magnetic field in our case, on Fig.~\ref{fig:fieldpot}.a. The effect is well approximated by a gradient $C' = 0.267 \, \mu\mathrm{m}^{-1}$ of the $\bm{C}\left(\bm{r}\right) = (C'y, 0, 0)$ vector. Figure~\ref{fig:fieldpot}.c presents the different potentials of the five $m_F$ sub-levels along the $y$-axis in our trap by choosing the quantization axis of angular momentum as the unit vector of $x$-axis. At first order, the traps are displaced with state-dependent trap centers: 
\begin{equation}
y_c \left(m_F\right) = \frac{w_0^2}{4} \frac{\delta_1 - \delta_2}{2\delta_1 + \delta_2} C' g_F m_F \simeq - m_F \times 16.7\, \mathrm{nm} \ .  
\label{eq:Udippoly0}
\end{equation}
We can see on Fig.~\ref{fig:fieldpot}.a that an atom crossing the $y$-axis while maintaining its spin will see a flip of the orientation of the fictitious magnetic field.  This situation is analogous to the evolution of atom spins crossing the center of a linear magnetic trap. In our linearly polarized dipole trap, the scalar potential provides an additional confinement which leads only to a displacement of the trap centers instead than spin flip losses.

If a large circular component is present in the polarization of the dipole light before the focus, it will induce a strong uniform $z$-component of the $\bm{C}$ vector, which will dominate over the linear polarization gradients, making them negligible. The dipole potential  is then given by :
\begin{equation}
U_{\mathrm{dip}}\left(\bm{r},m_F\right) = U_{\mathrm{dip},0}\left(\bm{r}\right) \left( 1 + \frac{\delta_1 - \delta_2}{2\delta_1 + \delta_2} g_F m_F \Vert\bm{C}\Vert\left(\bm{r}\right)   \right) 
\label{eq:Udipellip}
\end{equation}
where the quantization axis of the angular momentum is chosen in the
direction of the local $\bm{C}$ vector. The traps of the different
sub-levels have the same center, but differ in trap depth and oscillation frequencies, as
shown by Fig.~\ref{fig:fieldpot}.d. This effect increases with the
weight of the circular component of the polarization.  As can be seen
in Fig.~\ref{fig:fieldpot}.b, an atom crossing the $y$-axis will
experience a continuous evolution of the orientation of the fictitious
magnetic field whose variation decreases when the circular component
increases. For a large enough circular component, the atom spin will
be able to follow adiabatically the orientation of the field.

The natural question is then to determine what is the minimal amount of circular component (i.e. the smallest value of the $z$-component $C_z$ of the $\bm{C}$ vector) one needs to add to obtain an adiabatic evolution and thus mitigate the polarization gradient effects.
We can estimate a lower bound for $C_z$ by using the criterion for adiabatic evolution of the spin given by $\omega_L \gg \left|\mathrm{d}\theta/ \mathrm{d}t \right|$ where $\omega_L$ is the Larmor precession pulsation and $\left|\mathrm{d}\theta/ \mathrm{d}t \right|$ is the change of the orientation angle $\theta$ of the field during the atom motion. 
Considering that $\bm{C} = (C' y, 0, C_z)$ define the angle $\theta$ and using an harmonic approximation for the trap, one finds that the maximum of angle variation is obviously in the center of the trap where the atom speed $v_y$ is maximum and is given by  
$\left| \mathrm{d}\theta / \mathrm{d}t \right| = \left|C'/C_z\right| \left|v_y\right|$. The Larmor precession pulsation admit a minimum in the middle of the trap with value $\omega_L = \frac{U_{\mathrm{dip},0}\left(\bm{0}\right)}{\hbar} g_F \frac{\delta_1 - \delta_2}{2 \delta_1 + \delta_2} |C_z|$.
The evolution thus becomes adiabatic when the circular component fulfills:
\begin{equation}
|C_z| \gg \sqrt{ \frac{\hbar |C'| |v_y|  }{ U_{\mathrm{dip},0}\left(\bm{0}\right) g_F \frac{\delta_1 - \delta_2}{2 \delta_1 + \delta_2}} }. 
\end{equation} 
For our optical tweezer, we find that $|C_z| \gg 0.018 $ by estimating the maximum speed  $v_y$ at the center of an harmonic trap with by $v_y = \sqrt{2 k_B T_D/m_{\mathrm{Rb}}}$, where  $m_{\mathrm{Rb}}$ is the $^{87} Rb$ atom mass and $T_{\mathrm{D}} \simeq 140~\mu$K  the Doppler temperature. We find indeed experimentally that for our optimal polarization $|C_z| \simeq 0.243$. This is one order of magnitude larger than the estimated lower bound, ensuring an adiabatic evolution of the spin, that  minimizes the effect of state-dependent trap centers induced by the polarization gradient.

\section{\label{sec:heating}Heating mechanisms}

\begin{figure}[t]
\includegraphics[width = 0.95\columnwidth]{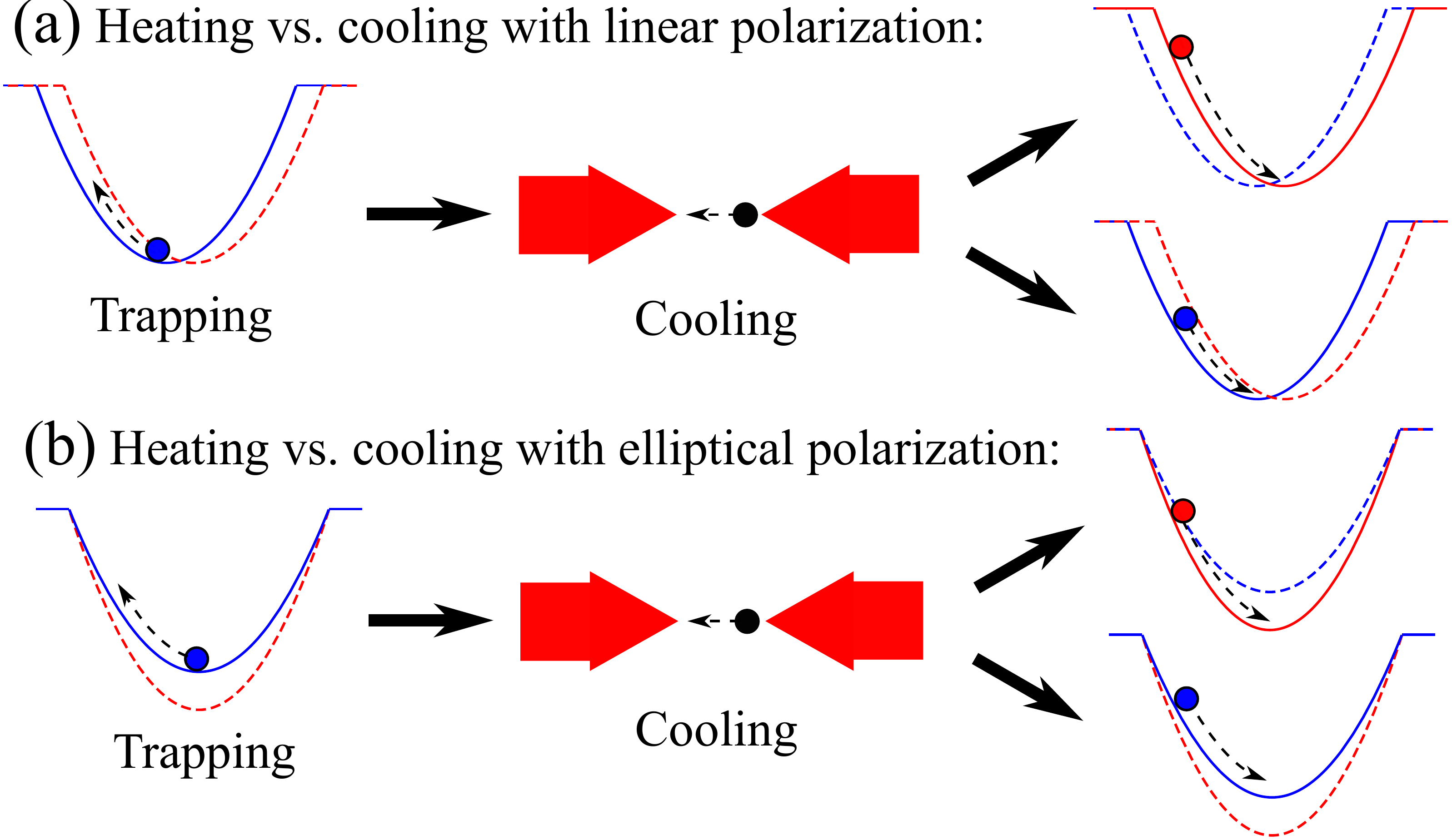}
\caption{Simplified schemes (with only two spin states represented in red and blue) of the cycles followed by the trapped atom. The MOT cooling when the dipole is off competes with the heating of trap potential fluctuations induced by the spin-state quantum jumps.}
\label{fig:heatscheme}
\end{figure} 
 
In order to get a better understanding, we identify different heating
processes of the trapped single atom that compete with the cooling
provided by the MOT beams.  For the linear and elliptic polarization
configurations, they originate from the change of spin sub-level $m_F$
due to the excitation by the MOT beams and desexcitation to the ground
state.  The evolution of an atom is schematically represented by
Fig.~\ref{fig:heatscheme}, for the linear and elliptic polarizations,
respectively. When the dipole trap is on, light shifts are so large
($117\,$MHz at the center) that the excitation by the MOT beams can be
neglected, and the atom oscillates freely in the trap while conserving
its spin state. When the dipole trap is switched off, the MOT beams
become almost resonant and provide both cooling and spin state
transitions. When the dipole laser is switched on again, the quantum
jumps of the spin-state may have led the atom to a different $m_F$-state
and so to experience a different trap. We have then three different heating mechanisms : 
\begin{itemize}
\item for any polarization: when the dipole trap is off, the atom can escape by spatial diffusion. This process is negligible if the chopping frequency is much larger than twice the largest trap
oscillation frequency.
\item for linear polarization: the spin-state diffusion gives rise to a random trap shaking due to the displacement of the state-dependent trap centers.
\item for elliptical polarization: the diffusion leads to heating by a random modulation of the trap frequencies.
\end{itemize}

The different heating effects are larger for lower chopping
frequencies, because the atom evolves longer in the trap and in the
MOT between two changes, which gives larger energy differences. An
optimal elliptic polarization exists mitigating the heating by random
trap shaking in the linearly polarized beam and the heating by random
trap-frequency modulation for moderate elliptic polarization, while
the modulation amplitude of the latter increases with the weight of
the circular component.

\section{\label{sec:simulation}Simulation of single atom lifetime}

To confirm the above explanation of the observed lifetime evolution with the trap modulation frequency in terms of trap modifications implied by quantum jumps of spin state, we perfom Monte-Carlo simulations of the atom motion in our pulsed trap. For a given polarization, the simulation considers the evolution of several thousands of atoms subjected to about a million cycles of alternating trapping and cooling phases. In order to obtain realistic computational times, we make some approximations that simplify the calculations while keeping the main physical features of the experimental situation. 

The first approximation consists in limiting the simulation to two dimensions given by the strong axes of the trap (transverse to the dipole laser propagation axis) which have oscillation frequencies an order of magnitude higher than the weak axis one. Thus, the energy change during one modulation cycle is much more important along these axes. We also consider truncated harmonic traps which provide a good approximation of the Gaussian potential in the center where the atom mostly evolves, and which have a depth limited by the maximum dipole light shift. For a linear polarization, the state dependent traps are shifted along the $y$-axis by $y_c \left(m_F\right)$ (cf. Eq~\ref{eq:Udippoly0}). For an elliptic polarization, we change the trap depth and frequencies according to Eq.~\ref{eq:Udipellip}.

The exact treatment of the cooling in $6$ nearly resonant MOT beams
implies to know the $5$ relative phases between the beams and to
resolve the $78$ coupled optical Bloch equations. We simplify this
description by taking an usual classical damping force
$- m_{\mathrm{Rb}} \gamma_v \bm{v}$ where $\gamma_v$ is the
characteristic cooling coefficient. In our setup, the cooling on the
$y$-axis is stronger. Due to geometrical hindrance, two pairs
of MOT beams have indeed a $10 \,$ deg deviation from this axis, so
we use here a cooling coefficient
$2 \cos(10 \, \mathrm{deg}) \gamma_v$.  The Doppler cooling limit is
considered by canceling this force if the atom kinetic energy is lower
than the thermal energy $k_{\mathrm{B}} T_{\mathrm{D}}$ at the Doppler
temperature when the cooling phase starts.  To approximately describe
the change in $m_F$ spin state of the atom, we calculate, from the
Clebsch-Gordan coefficients, the probabilities of excitation to the
different sub-levels of the $5^2P_{3/2} \ F'=3$ excited state in an
isotropic light field and the probabilities of spontaneous emission
down to the $5^2S_{1/2} \ F=2$ sub-levels. It gives us the
probabilities to change from a state to another per photon scattering
event. For a given initial state, the final state probabilities after
the cooling phase are then given by its duration
$\tau_{1/2} = 1/(2f_{\mathrm{m}})$, the average scattering rate
$\Gamma_{\mathrm{ch}}$ and the analytic solution of the coupled
equations of state population evolution. These two parameters
$\gamma_v$ and $\Gamma_{\mathrm{ch}}$ of the cooling phase are
difficult to estimate carefully in our system because we need intense
MOT beams ($50~\mathrm{mW}/\mathrm{cm}^2$) to provide efficient trap
loading, thus we can not apply the usual low saturation
limit. Nevertheless, an heuristic equation established in
Ref.~\cite{wohlleben2001atom}~(Eq. 4) and the measured rate of
spontaneously emitted photons in the detection respectively give some
approximate initial values of $\gamma_v$ and
$\Gamma_{\mathrm{ch}}$. These values are then adjusted to fit the
experimental data. We find $\gamma_v = 3.3\,$kHz which is $30\%$ lower
that the initial value, and with $\Gamma_{\mathrm{ch}} = 35\,$MHz that
is roughly three time higher than the observed spontaneous
emission. The latter result can be explained by the fact that we are
using beam intensities well above the saturation, where stimulated
emission processes dominate.

\begin{figure}[tb]
\includegraphics[width = 0.95\columnwidth]{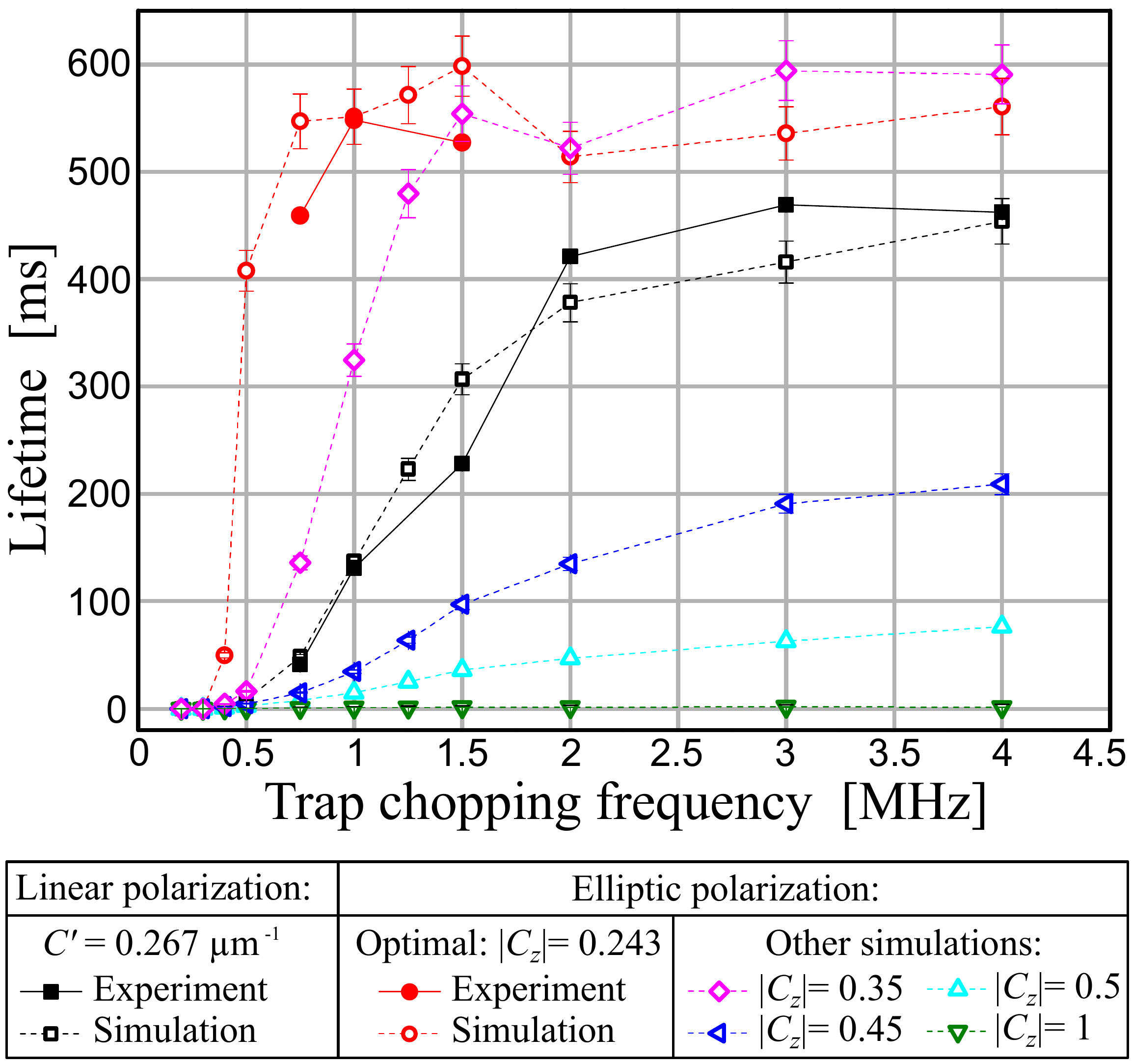}
\caption{Results of lifetime as function of trap pulsing frequency given by atomic motion simulations for different dipole trap light polarization with $500$ simulated atoms for each point. The error bars represent the $4.7\%$ relative standard deviation for $500$ events of an exponential distribution. }
\label{fig:sim}
\end{figure}

We initiate a simulation by picking an atom in $m_F = 0$ state and randomly distributed in phase space with a Gaussian density probability characterized by the  Doppler temperature $T_{\mathrm{D}}$ and the trap frequency. We start the first cycle by the trapping phase and calculate the position and speed after a chopping period half $\tau_{1/2}$ of evolution. Then, we let the atom move under the cooling force for another $\tau_{1/2}$. Afterward, we randomly change the internal state according to the calculated probabilities and reckon the atom energy from its state, position and speed in the dipole trap. We repeat this cycle until the atom energy reaches a value higher than the trap depth (which means that it escapes from the trap) or until the time exceeds a random value from an exponential distribution. The latter takes into account the background gas collisions which limit the atom lifetime to $550\,$ms according to the experimental data.

By using the simulation without spin state changes, we first verified that  the heating due to the chopping of the dipole trap is negligible for chopping frequencies higher than $400\,$kHz, as expected since $2 f_{\mathrm{t}} \simeq 340 \,$kHz. For the different polarization configurations in presence of spin-state diffusion, the lifetimes extracted from the survival probabilities of $500$ simulated atoms are presented on Fig.~\ref{fig:sim}. The results show the improvement of the lifetime that was observed experimentally for an elliptic polarization with a small circular component compared to a linear polarization, and agree quantitatively well with the experimental data. As a larger circular polarization component of the dipole light is added ($\vert C_z \vert$ increasing), the lifetime decreases and almost cancels for a pure circular polarization even for large modulation frequencies. This explains why we were not able to detect any atom for large circular components in the polarization of the dipole light. These simulations confirm the existence of an optimal elliptic polarization that limits heating due to atom spin state quantum jumps in high NA tweezers.

\section{\label{sec:conclusion}Conclusion and Outlook}
We have shown how some deleterious effects of polarization gradients
in strongly focused optical tweezers can be mitigated by using
elliptical polarization of the incoming dipole light. 
This method is general and can be applied in any experiment where the
light field is tightly confined so that  the longitudinal component of
the polarization is non-negligible. Besides strongly focused optical
tweezers, this also includes optical nanofiber \cite{Mitsch2014chiralfiber} and optical microcavities \cite{Junge2013WGM,Shomroni2014WGM}.

\begin{acknowledgments}
We acknowledge funding from the ANR SAROCEMA project, grant ANR-14-CE32-0002 of the French Agence Nationale de la Recherche and from \'{E}mergence-UPMC-2009 research program. We thank D.~Maxein and L.~Hohmann for contributions in the
early stage of the experiment.
\end{acknowledgments}


%


\end{document}